\documentstyle[12pt,a41,epsfig]{article}
\newcommand{\bq}{\begin{equation}}
\newcommand{\eq}{\end{equation}}

\newcommand{\gsim}{\raisebox{-0.07cm}{$\, \stackrel{>}{{\scriptstyle
\sim}}\, $}}

\newcommand\GeV{\,\mbox{GeV}}
\newcommand\TeV{\,\mbox{TeV}}

\begin{document}
\sloppy
\thispagestyle{empty}
\begin{flushleft}
DESY 97--067  \hfill
{\tt hep-ph/9811271}\\
December 1997
\end{flushleft}

\mbox{}
\vspace*{\fill}
\begin{center}
{\LARGE\bf  Leptoquark Pair Production Cross Sections at} \\

\vspace{2mm}
{\LARGE\bf  Hadron Colliders
}\\

\vspace{2em}
\large
Johannes Bl\"umlein$^a$, Edward Boos$^{a,b}$ and 
Alexander Kryukov$^{a,b}$
\\
\vspace{2em}
{\it $^a$DESY -- Zeuthen,}
 \\
{\it Platanenallee 6,
D--15735 Zeuthen, Germany}\\

\vspace{3mm}
{\it $^b$Institute of Nuclear Physics}\\
{\it
Moscow State University, RU--119899
Moscow, Russia} \\
\end{center}
\vspace*{\fill}
\begin{abstract}
\noindent
A compilation is given of the pair production cross sections for scalar
and vector leptoquarks in the kinematic range of the hadron colliders
Tevatron and LHC.
\end{abstract}
\vspace*{\fill}
\newpage
\noindent
%%%%%%%%%%%%%%%%%%%%%%%%%%%%%%%%%%%%%%%%%%%%%%%%%%%%%%%%%%%%%%%%%%%%%%%%
\section{Introduction}
%%%%%%%%%%%%%%%%%%%%%%%%%%%%%%%%%%%%%%%%%%%%%%%%%%%%%%%%%%%%%%%%%%%%%%%%

\vspace{1mm}
\noindent
In many extensions of the Standard Model leptoquark  states do emerge.
The couplings of the leptoquarks to the gauge sector are predicted due
to the gauge symmetries, up to eventual anomalous couplings in the case 
of vector leptoquarks, whereas the fermionic couplings $\lambda$ are free
parameters of the models. Model--independent searches do therefore refer 
to the pair--production processes~\cite{BBK1}, which are widely dominated
by the gauge-couplings of leptoquarks. Future searches may be carried out
both at the forthcoming hadron and linear $e^+e^-$ colliders. The 
production cross sections are largest at hadron colliders in general.
Specific channels, on the other hand, may be identified unambiguously 
only in $e^+e^-$ annihilation~\cite{BR1,BBK2}.

In this note a compilation of the hadronic pair--production cross
sections for leptoquarks at hadron colliders is provided for the kinematic 
range of Tevatron and LHC. The scattering cross sections were calculated 
by the code {\tt LQPAIR 1.00}~\cite{LQPAIR} both for scalar and vector 
leptoquarks. In the latter case typical choices for the anomalous 
couplings $\kappa_G$ and $\lambda_G$ to the gluon are considered. For 
scalar leptoquarks also the $O(\alpha_s)$-corrections~\cite{KRAE} are 
presented. The numerical values given below correspond to the integrated 
cross section values without cuts and may be used in the experimental 
analyses to obtain fast estimates in the forthcoming searches. This is 
particularly useful for the case of vector leptoquarks for which time 
consuming minimizations have to be carried out in searching the complete 
parameter space.
%%%%%%%%%%%%%%%%%%%%%%%%%%%%%%%%%%%%%%%%%%%%%%%%%%%%%%%%%%%%%%%%%%%%%%%%
\section{Scattering Cross Sections}
%%%%%%%%%%%%%%%%%%%%%%%%%%%%%%%%%%%%%%%%%%%%%%%%%%%%%%%%%%%%%%%%%%%%%%%%

\vspace{1mm}
\noindent
Leptoquarks may be searched for at hadron colliders both studying
single and pair production processes. In the case of the single 
production  processes~\cite{ZER} the reaction cross sections are
$\propto  \lambda^2$ and amount to
$\sigma_{\rm sing} \sim 0.4~fb ... 1.3~fb$
at Tevatron energies for  fermionic couplings of the order
$(\lambda/e)\sqrt{Br}  \approx 0.075$,
only, [1a], and are too small to be detected currently. These values
of $\lambda$ would correspond to an interpretation of the HERA-events
as due to single leptoquark-production, setting an upper bound.

On the other hand, given the small fermionic couplings, the pair
production processes~\cite{BBK1} depend on
the leptoquark--gluon couplings only. In the case of scalar leptoquarks
the production cross section is completely predicted, whereas it
depends on anomalous couplings, such as  $\kappa_G$ and $\lambda_G$, in
the case of vector leptoquarks. As was shown in ref.~\cite{BBK1},
however, there exists a global minimum ${\sf min}_{\kappa_G, \lambda_G}
\left [\sigma(\Phi_V \overline{\Phi}_V)\right]  > 0$
allowing for a model-independent analysis.

The production cross sections for scalar  leptoquarks in the
partonic subsystem read~\cite{BBK1}
%-----------------------------------------------------------------------
\begin{eqnarray}
\sigma_{\Phi_S \overline{\Phi}_S}^{q \overline{q}} &=&
\frac{2 \pi \alpha_s^2}{27 \hat{s}} \beta^3\\
\sigma_{\Phi_S \overline{\Phi}_S}^{gg} &=&
 \frac{\pi \alpha_s^2}{
96 \hat{s}} \left [ \beta (41 - 31 \beta^2) - (17 - 18 \beta^2 + \beta^4)
\ln \left|\frac{1+\beta}{1-\beta}\right| \right],
\end{eqnarray}
%-----------------------------------------------------------------------
with $\beta = \sqrt{1 - 4 M_{\Phi}^2/\hat{s}}$. The $O(\alpha_s)$
correction  to the production cross section was calculated in 
ref.~\cite{KRAE} and amounts to a $K$-factor of 1.12 only for the choice 
$\mu = M_{\Phi}$ of  the factorization scale, for $M_{\Phi} \simeq
200 \GeV$. The cross sections in the case of vector leptoquark pair 
production are more complicated, cf.~\cite{BBK1}, due to the presence of 
the anomalous couplings $\kappa_G$ and $\lambda_G$ and have the general 
structure
%-----------------------------------------------------------------------
\begin{eqnarray}
\sigma_{\Phi_V \overline{\Phi}_V}^{q \overline{q}} &=&
\frac{4 \pi \alpha_s^2}{9 M_V} \sum_{i=0}^5 \chi_i^q(\kappa_G,\lambda_G)
\widetilde{G}_i(\hat{s}, \beta)\\
\sigma_{\Phi_V \overline{\Phi}_V}^{gg} &=&
\frac{\pi \alpha_s^2}{96 M_V} \sum_{i=0}^{14}
\chi_i^g(\kappa_G,\lambda_G)
\widetilde{F}_i(\hat{s}, \beta)~.
\end{eqnarray}
%-----------------------------------------------------------------------
The functions $\chi^{q,g},\widetilde{G}_i$ and $\widetilde{F}_i$ are
given in ref.~\cite{BBK1}.
For $\kappa_G = \lambda_G = 0$ (Yang--Mills type couplings)
one obtains~\cite{BBK1}
%-----------------------------------------------------------------------
\begin{eqnarray}
\sigma_{\Phi_V \overline{\Phi}_V}^{q \overline{q}} =
\frac{ \pi \alpha_s^2}{54 M_V} \left[ \frac{\hat{s}}{M_V^2}
+ 23 - 3 \beta^2 \right],
\sigma_{\Phi_V \overline{\Phi}_V}^{gg} =
 \frac{\pi \alpha_s^2}{96 M_V} \left\{\beta
A(\beta) - B(\beta)
\ln \left|\frac{1+\beta}{1-\beta}\right| \right \},
\nonumber
\end{eqnarray}
%-----------------------------------------------------------------------
\begin{eqnarray}
A(\beta) = \frac{523}{4} - 90 \beta^2 + \frac{93}{4} \beta^4,~~~~~
B(\beta) = \frac{3}{4} \left[
65 - 83 \beta^2 + 19 \beta^4 - \beta^6 \right]~.
\end{eqnarray}
%-----------------------------------------------------------------------
Choosing the factorization and renormalization scales by
$\mu = M_{\Phi}$
the pair production cross sections   for scalar and vector
leptoquarks (minimizing for $\kappa_G$ and $\lambda_G$)
at Born level
and using the parametrization~\cite{CTEQ}
for the parton densities are~[1a]~:
%-----------------------------------------------------------------------
\begin{eqnarray}
\label{J:E:CS}
\sigma_S(M_{\Phi} = 200 \GeV) = 0.16~pb~~~~~~~\sigma_V(M_{\Phi} =
200 \GeV) = 0.29~pb.
\end{eqnarray}
%-----------------------------------------------------------------------

%%%%%%%%%%%%%%%%%%%%%%%%%%%%%%%%%%%%%%%%%%%%%%%%%%%%%%%%%%%%%%%%%%%%%%%%
\section{Numerical values of  the Integral Cross Sections}
%%%%%%%%%%%%%%%%%%%%%%%%%%%%%%%%%%%%%%%%%%%%%%%%%%%%%%%%%%%%%%%%%%%%%%%%

\vspace{2mm}
\noindent
%-----------------------------------------------------------------------
The mass-range for first-generation scalar leptoquarks already excluded
by the Tevatron experiments is
%-----------------------------------------------------------------------
\begin{eqnarray}
\label{J:E:TL}
 M  <~~~~~~~213~\GeV & & {\rm~CDF}~~~Br(eq) = 1\nonumber\\
 M  < 176~(225) \GeV & & {\rm~D0}~~~~~~Br(eq) = 0.5~(1) \\
 M  <~~~~~~~242~\GeV & & {\rm~combined}~~~Br(eq) = 1\nonumber
\end{eqnarray}
%-----------------------------------------------------------------------
at 95\%~CL~\cite{SLIM}.
The mass bounds for vector leptoquarks are correspondingly higher because
of the larger production cross section. These limits have still to be 
determined by the Tevatron experiments. The region for future searches is
thus $M \gsim 200 \GeV$, because the branching ratios $Br(eq)$ can be as 
small as 0.5.

In Figures~1,2 and 5 the scalar leptoquark pair--production cross sections
are shown for cms energies of $\sqrt{S} = 1.8$ and $2 \TeV$ (Tevatron)
and $14 \TeV$ (LHC) as a function of the leptoquark mass choosing the 
factorization and renormalization scales $\mu = M_{\Phi}$. The band due 
to a variation of this scale in the range 
$\mu~\epsilon~[M_{\Phi}/2,2 M_{\Phi}]$ is also given. The cross sections
both for scalar and vector leptoquarks behave almost as
%----------------------------------------------------------------------
\begin{equation}
\log(\sigma) \sim A - B M_{\Phi}
\end{equation}
%----------------------------------------------------------------------
in the region of large masses.

Figures~3 and 5 present the mass-dependence of the $O(\alpha_s)$ 
$K$-factor at Tevatron and LHC, respectively, with values between
1.05 and 1.15 for $200 < M_{\Phi} < 350 \GeV$ for Tevatron and
1.35 to 1.55 for $200 < M_{\Phi} <1500 \GeV$ for LHC.

In Figures~6--11 the integrated cross sections for vector leptoquark
pair--production are shown correspondingly. Three typical cases are
considered for the anomalous couplings:
%----------------------------------------------------------------------
\begin{itemize}
\item
Yang-Mills type coupling (YM)~:   $\kappa_G = \lambda_G = 0$
\item
minimal coupling (MC)~:   $\kappa_G = 1,\lambda_G = 0$
\item
the global minimum of the cross section w.r.t. $\kappa_G$ and 
$\lambda_G$ for fixed $M_{\Phi}$~.
\end{itemize}
%----------------------------------------------------------------------

\noindent
The minimal cross sections are neither obtained for the Yang-Mills-type
or the minimal couplings, showing that a search in the two-parameter
space ($\kappa_G, \lambda_G$) is required to obtain a global bound.

\vspace*{2mm}
\noindent
{\bf Acknowledgement.}
We would like to thank M. Kr\"amer for useful discussions and making a
code for the calculation of the $K$-factor for the scalar
leptoquark production cross section available to us. E.B. and A.K. would
like to thank DESY-Zeuthen for hospitality. The work was supported in 
part by EU contract FMRX-CT98-0194(DG 12 - MIHT) and the RFBR grants 
96-02-19773a and 98-02-17699.

%%%%%%%%%%%%%%%%%%%%%%%%%%%%%%%%%%%%%%%%%%%%%%%%%%%%%%%%%%%%%%%%%%%%%%%%%%

\newpage
\begin{center}

\mbox{\epsfig{file=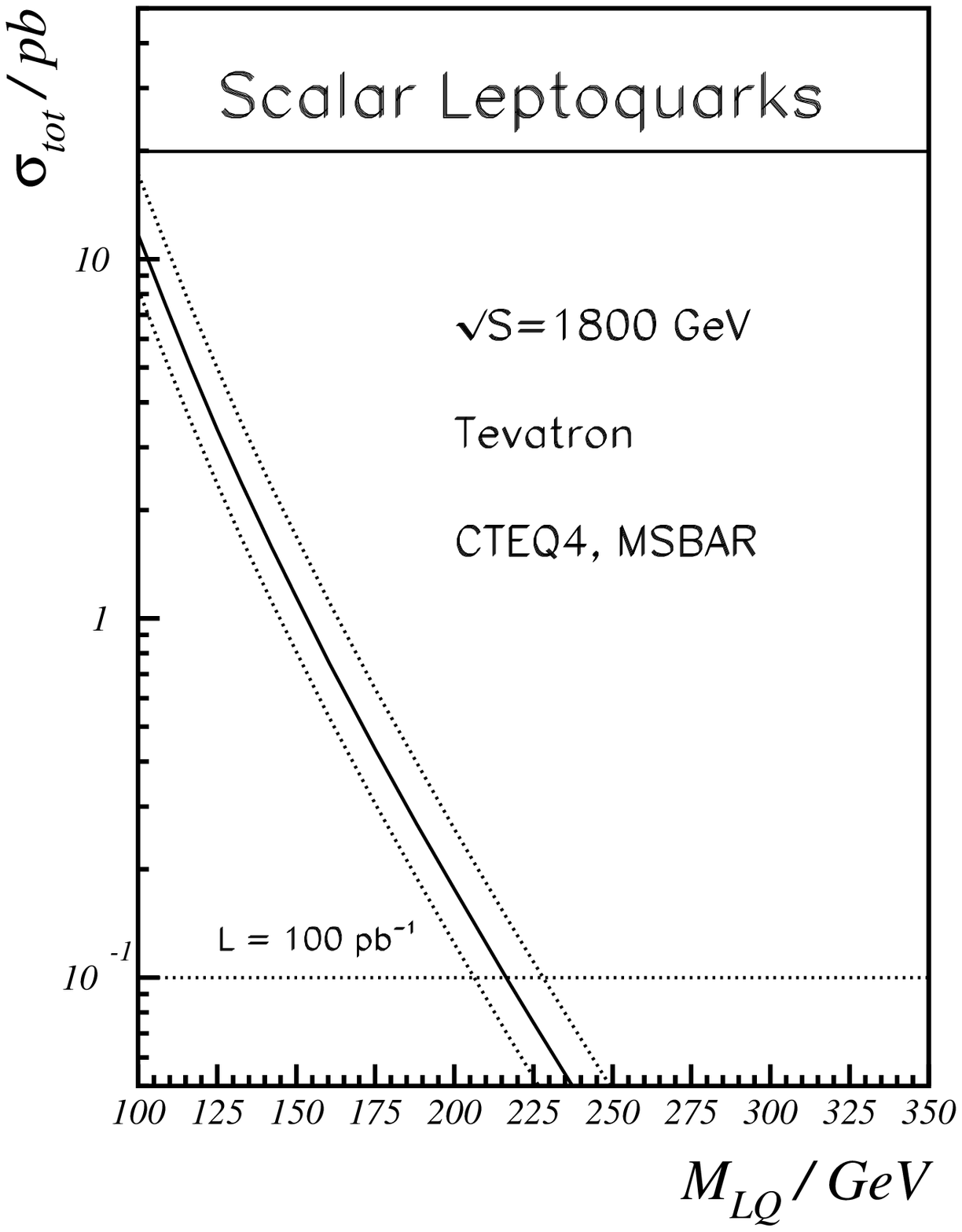,height=18cm,width=16cm}}

\vspace{2mm}
\noindent
\small
\end{center}
{\sf
Figure~1:~Integrated cross
sections for scalar leptoquark pair production at the
Tevatron,
$\sqrt{S}~=~1.8~\TeV$, $\mu = M_{\rm LQ}$, full line.
Dotted lines~: range of scale variation $\mu~\epsilon~[M_{\rm LQ}/2,
2 M_{\rm LQ}]$.}
\normalsize
%%%%%%%%%%%%%%%%%%%%%%%%%%%%%%%%%%%%%%%%%%%%%%%%%%%%%%%%%%%%%%%%%%%%%%%%%%

\newpage
\begin{center}

\mbox{\epsfig{file=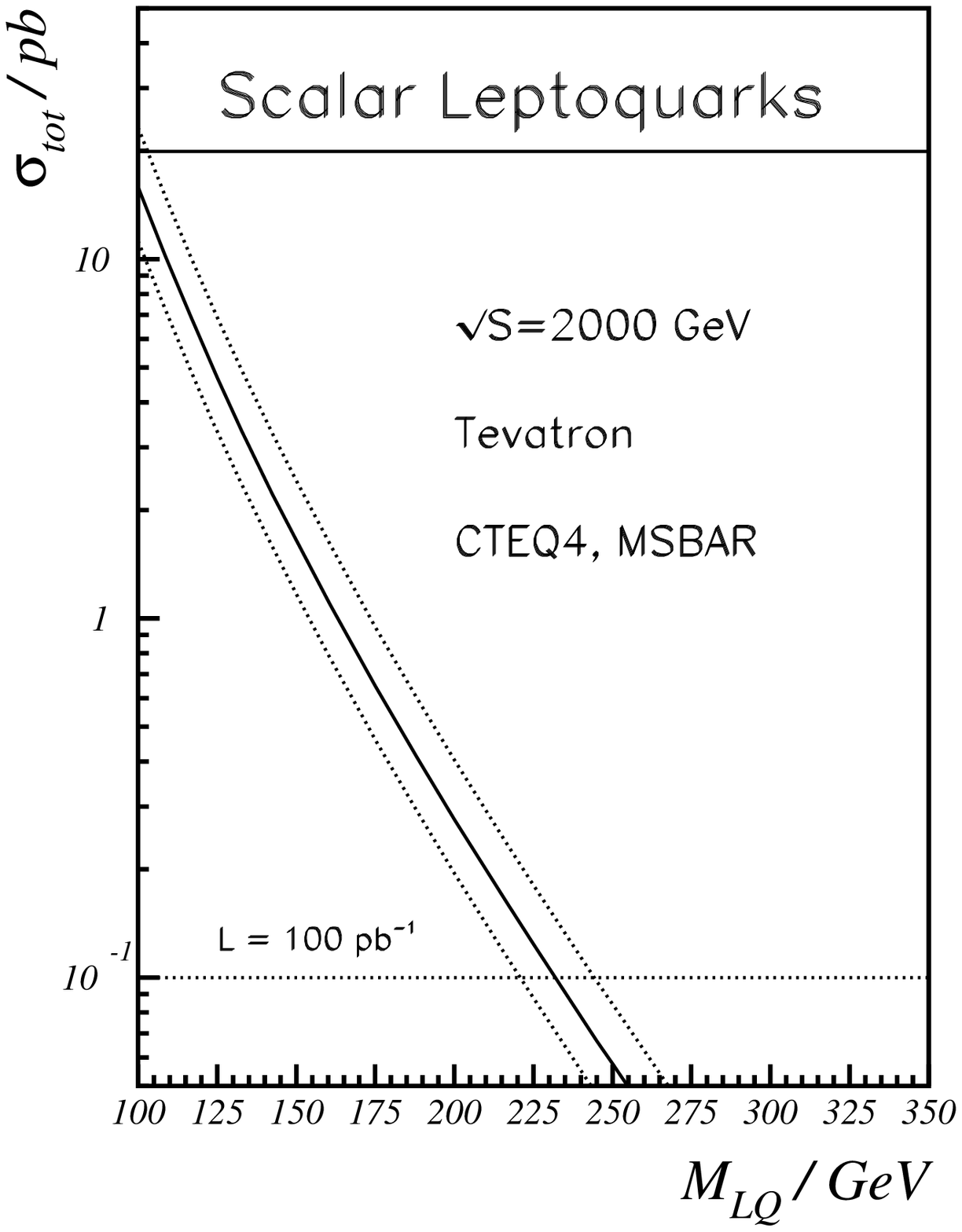,height=18cm,width=16cm}}

\vspace{2mm}
\noindent
\small
\end{center}
{\sf
Figure~2:~Integrated cross sections for scalar leptoquark pair
production at Tevatron at $\sqrt{S}~=~2.0~\TeV$.
Full line~: $\mu = M_{\rm LQ}$,
dotted lines~: range of scale variation $\mu~\epsilon~[M_{\rm LQ}/2,
2 M_{\rm LQ}]$.}
\normalsize
%%%%%%%%%%%%%%%%%%%%%%%%%%%%%%%%%%%%%%%%%%%%%%%%%%%%%%%%%%%%%%%%%%%%%%%%%%
\newpage
\begin{center}

\mbox{\epsfig{file=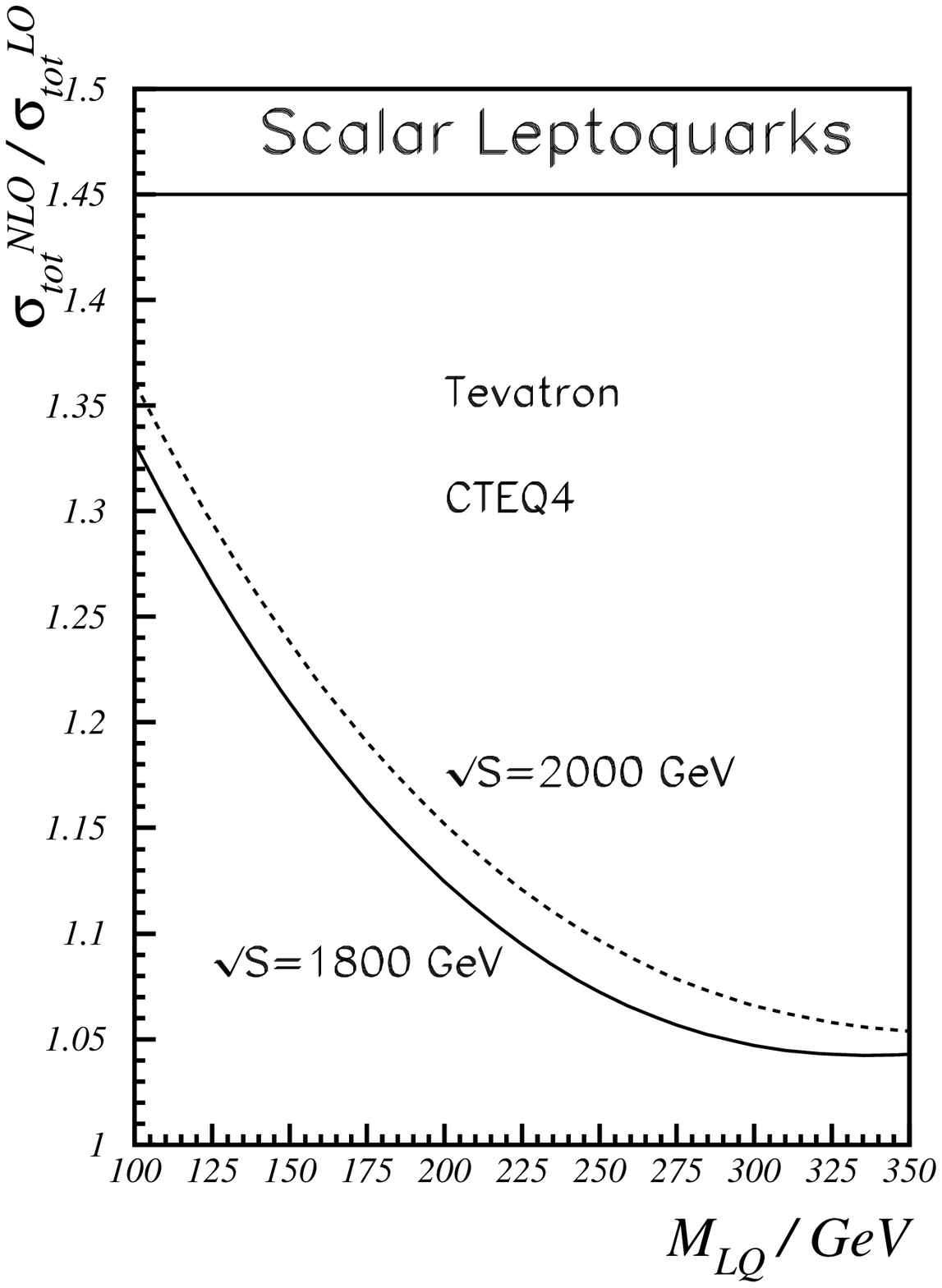,height=18cm,width=16cm}}

\vspace{2mm}
\noindent
\small
\end{center}
{\sf
Figure~3:~Ratio of the scalar pair production cross section in
next-to-leading and leading order QCD at Tevatron, $\mu = M_{\rm LQ}$.
Full line~:
$\sqrt{S}~=~1.8~\TeV$, dashed  line~: $\sqrt{S}~=~2.0~\TeV$.}
\normalsize
%%%%%%%%%%%%%%%%%%%%%%%%%%%%%%%%%%%%%%%%%%%%%%%%%%%%%%%%%%%%%%%%%%%%%%%%%%

\newpage
\begin{center}

\mbox{\epsfig{file=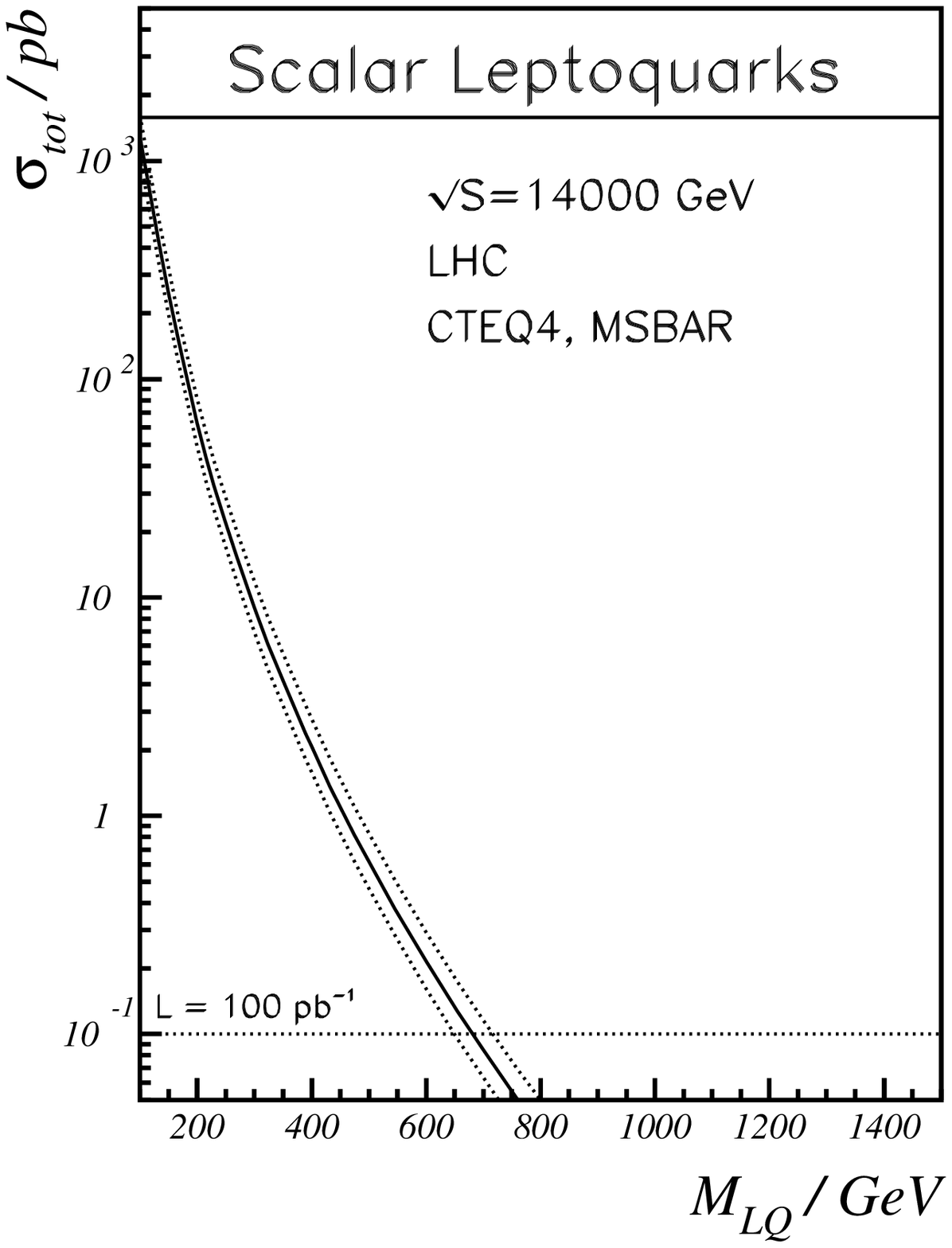,height=18cm,width=16cm}}

\vspace{2mm}
\noindent
\small
\end{center}
{\sf
Figure~4:~Integrated cross sections for scalar leptoquark pair
production at LHC at $\sqrt{S}~=~14~\TeV$. 
Full line~: $\mu = M_{\rm LQ}$,
dotted lines~: range of scale variation $\mu~\epsilon~[M_{\rm LQ}/2,
2 M_{\rm LQ}]$.}
\normalsize
%%%%%%%%%%%%%%%%%%%%%%%%%%%%%%%%%%%%%%%%%%%%%%%%%%%%%%%%%%%%%%%%%%%%%%%%%%
\newpage
\begin{center}

\mbox{\epsfig{file=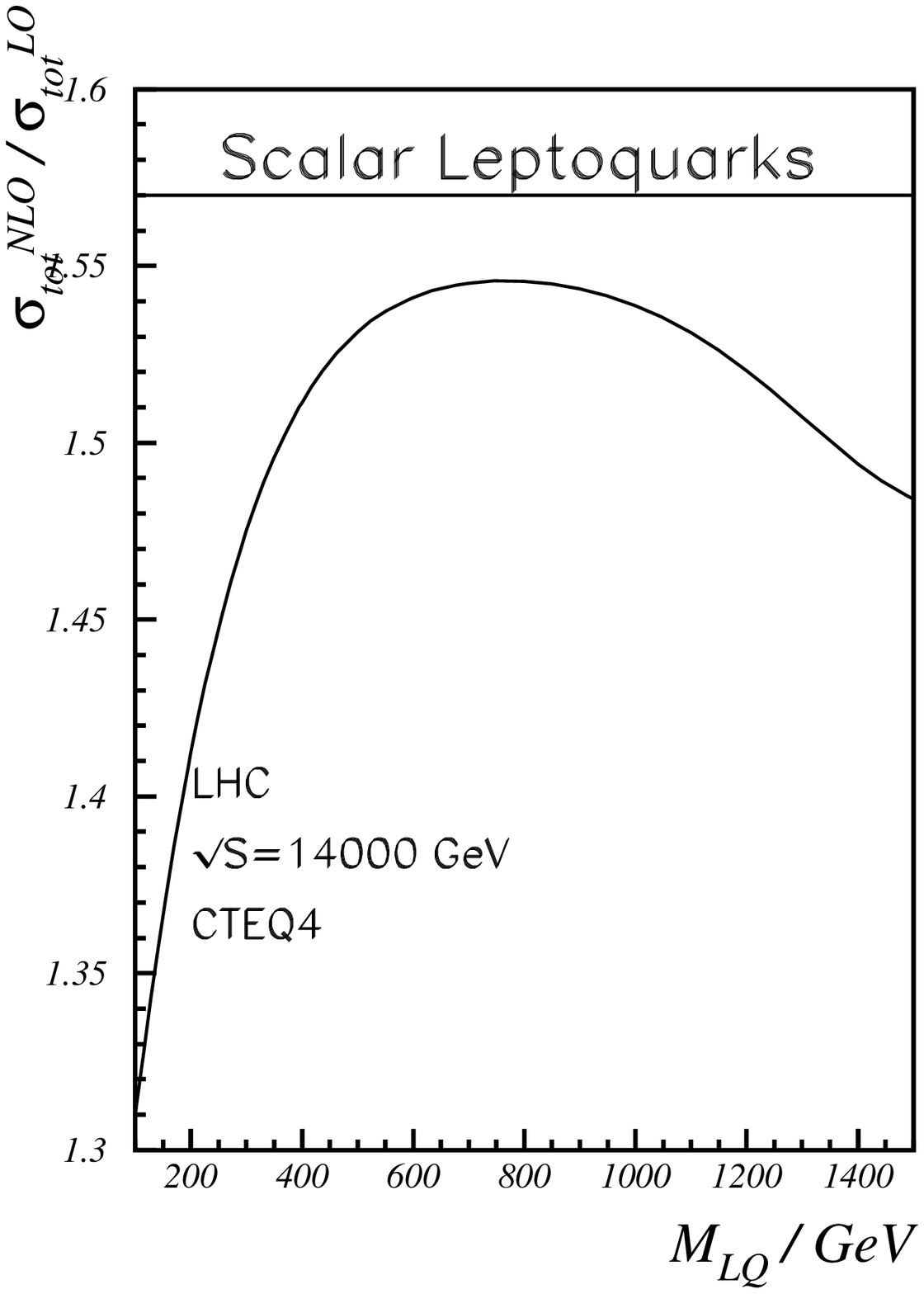,height=18cm,width=16cm}}

\vspace{2mm}
\noindent
\small
\end{center}
{\sf
Figure~5:~Cross section ratio
$\sigma_{\rm tot}^{\rm NLO}/\sigma_{\rm tot}^{\rm LO}$
for scalar leptoquark pair production at LHC,
$\sqrt{S}~=~14~\TeV$.}
\normalsize
%%%%%%%%%%%%%%%%%%%%%%%%%%%%%%%%%%%%%%%%%%%%%%%%%%%%%%%%%%%%%%%%%%%%%%%%%%
\newpage
\begin{center}

\mbox{\epsfig{file=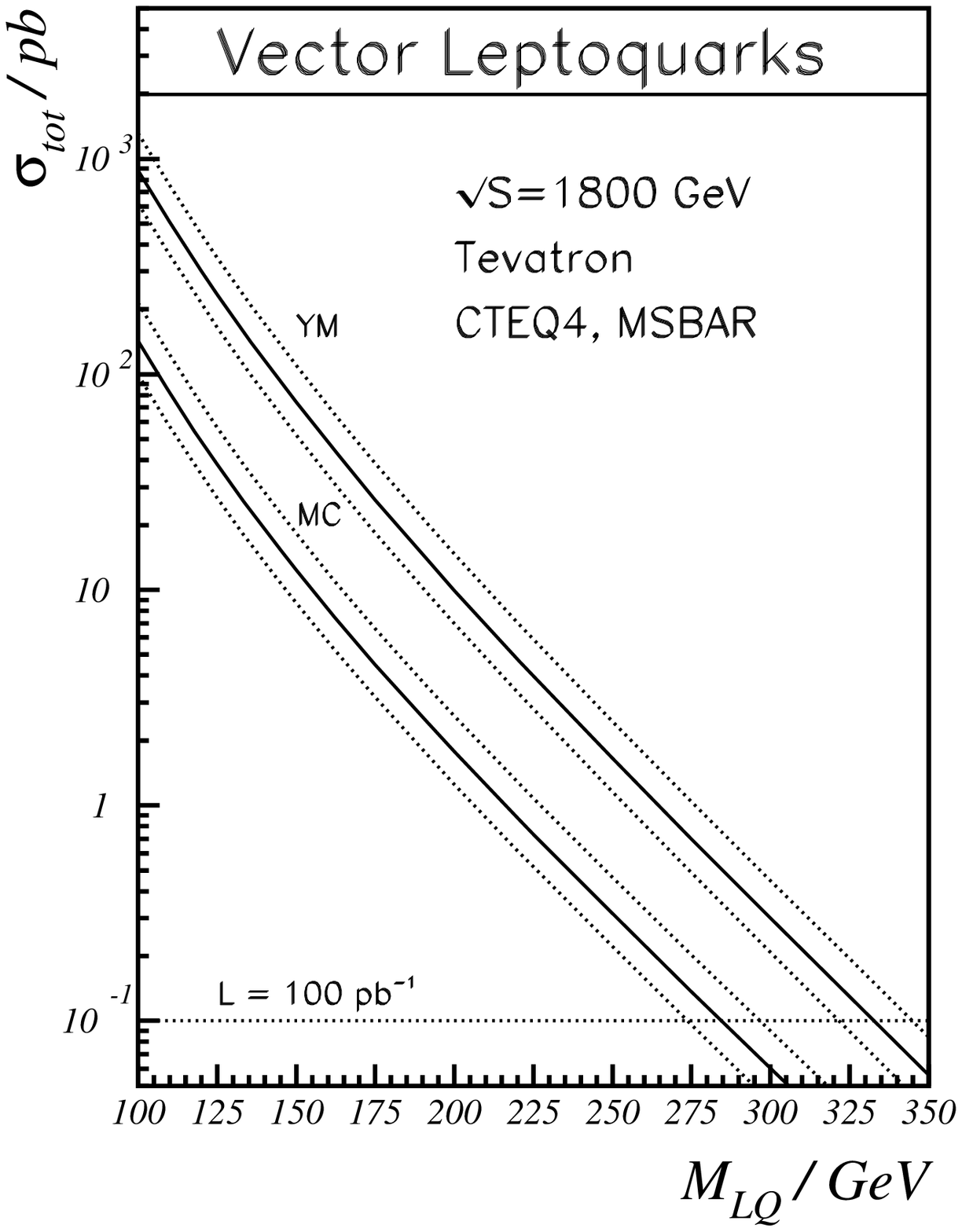,height=18cm,width=16cm}}

\vspace{2mm}
\noindent
\small
\end{center}
{\sf
Figure~6:~Integrated cross sections for vector leptoquark pair
production at Tevatron for the Yang--Mills type coupling (YM) and minimal
coupling (MC) at $\sqrt{S}~=~1.8~\TeV$. Full line~: $\mu = M_{\rm LQ}$,
dotted lines~: range of scale variation $\mu~\epsilon~[M_{\rm LQ}/2,
2 M_{\rm LQ}]$.}
\normalsize
%%%%%%%%%%%%%%%%%%%%%%%%%%%%%%%%%%%%%%%%%%%%%%%%%%%%%%%%%%%%%%%%%%%%%%%%%%

\newpage
\begin{center}

\mbox{\epsfig{file=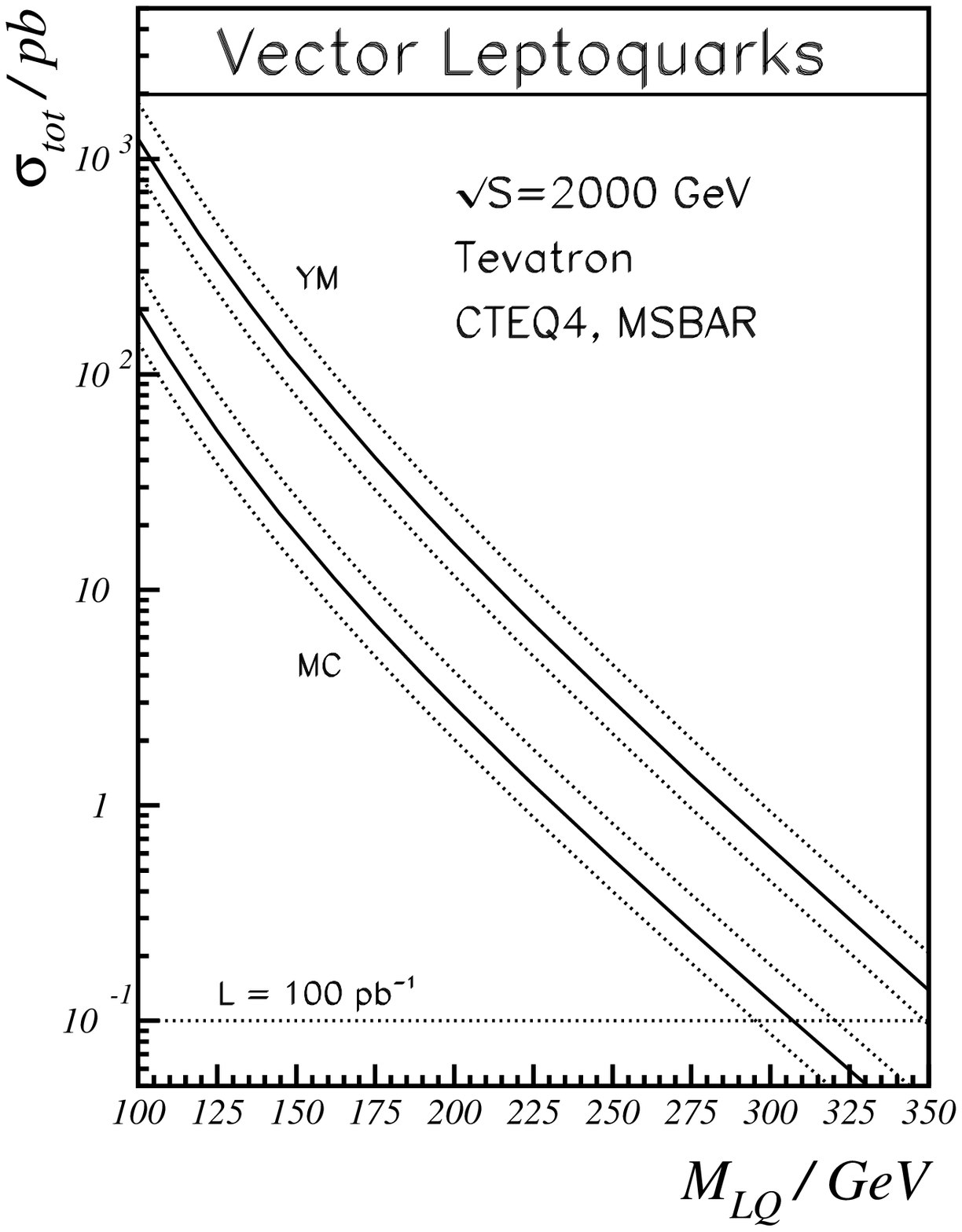,height=18cm,width=16cm}}

\vspace{2mm}
\noindent
\small
\end{center}
{\sf
Figure~7:~Integrated cross sections for vector leptoquark pair
production at Tevatron for the Yang--Mills type coupling (YM) and minimal
coupling (MC) at $\sqrt{S}~=~2.0~\TeV$. Full line~: $\mu = M_{\rm LQ}$,
dotted lines~: range of scale variation $\mu~\epsilon~[M_{\rm LQ}/2,
2 M_{\rm LQ}]$.}
\normalsize
%%%%%%%%%%%%%%%%%%%%%%%%%%%%%%%%%%%%%%%%%%%%%%%%%%%%%%%%%%%%%%%%%%%%%%%%%%
\newpage
\begin{center}

\mbox{\epsfig{file=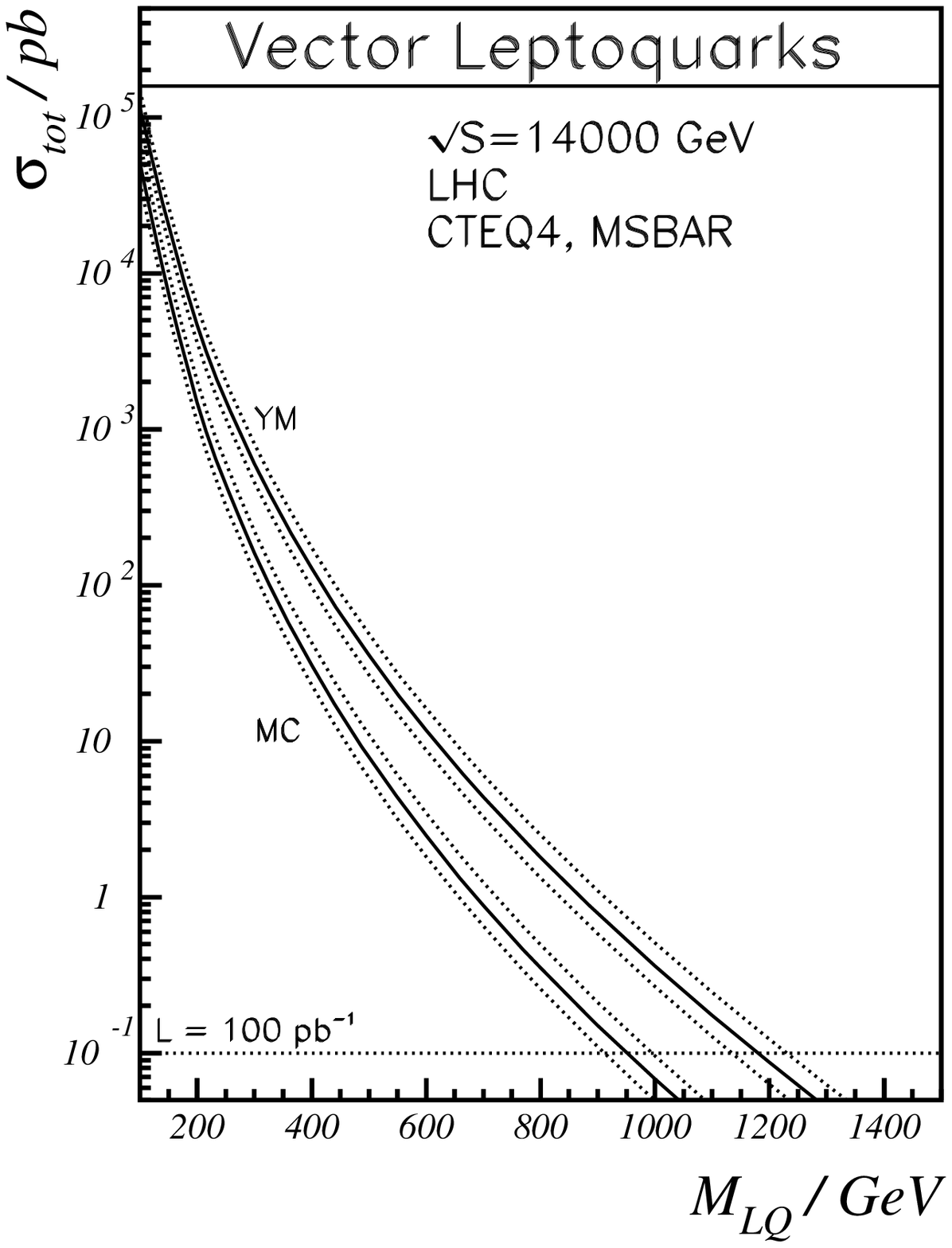,height=18cm,width=16cm}}

\vspace{2mm}
\noindent
\small
\end{center}
{\sf
Figure~8:~Integrated cross sections for vector leptoquark pair
production at LHC for the Yang--Mills type coupling (YM) and minimal
coupling (MC) at $\sqrt{S}~=~14~\TeV$. Full line~: $\mu = M_{\rm LQ}$,
dotted lines~: range of scale variation $\mu~\epsilon~[M_{\rm LQ}/2,
2 M_{\rm LQ}]$.}
\normalsize
%%%%%%%%%%%%%%%%%%%%%%%%%%%%%%%%%%%%%%%%%%%%%%%%%%%%%%%%%%%%%%%%%%%%%%%%%%

\newpage
\begin{center}

\mbox{\epsfig{file=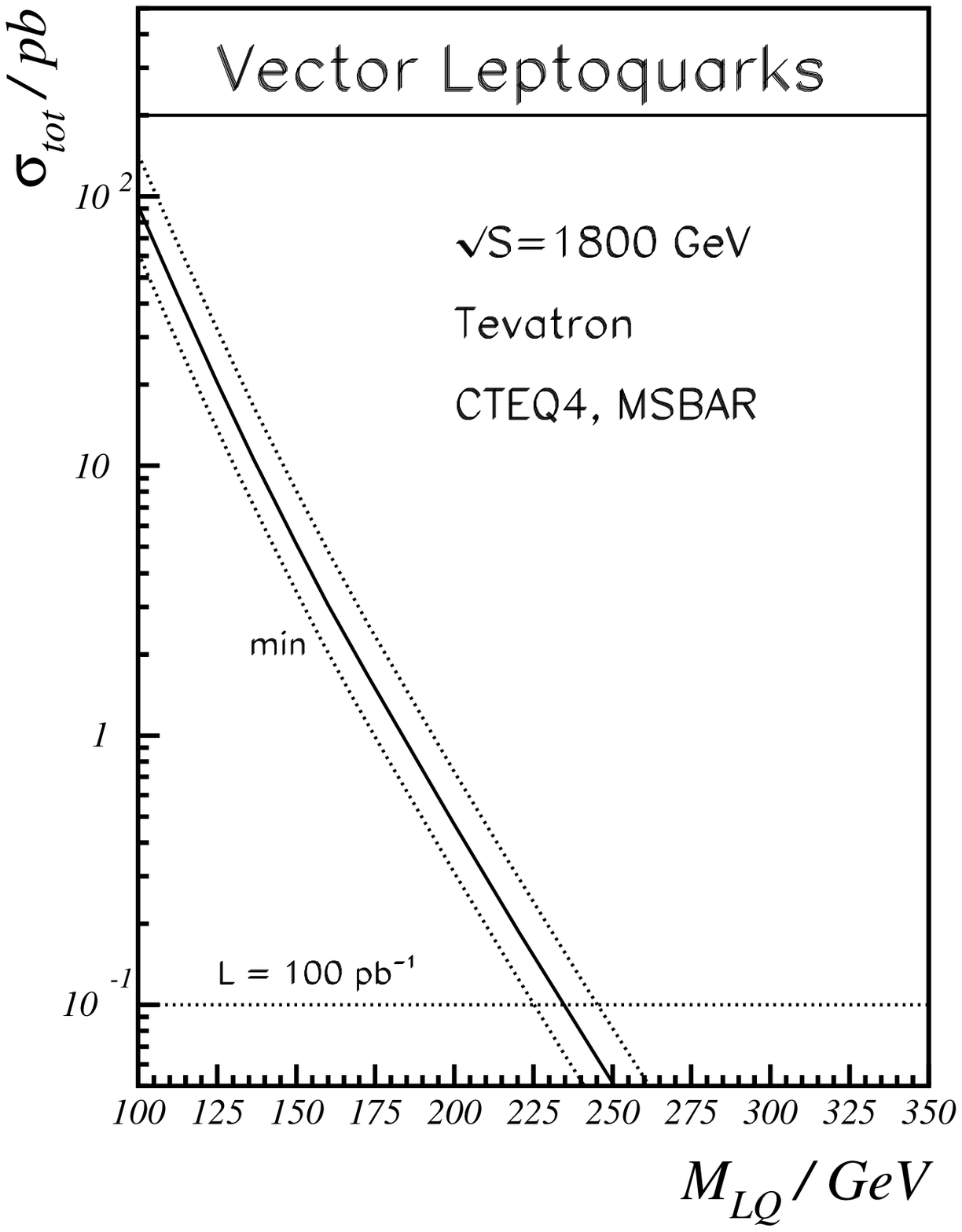,height=18cm,width=16cm}}

\vspace{2mm}
\noindent
\small
\end{center}
{\sf
Figure~9:~Integrated cross sections for vector leptoquark pair
production at Tevatron minimizing the cross section for $\kappa_G$ and
$\lambda_G$ at $\sqrt{S}~=~1.8~\TeV$. Full line~: $\mu = M_{\rm LQ}$,
dotted lines~: range of scale variation $\mu~\epsilon~[M_{\rm LQ}/2,
2 M_{\rm LQ}]$.}
\normalsize
%%%%%%%%%%%%%%%%%%%%%%%%%%%%%%%%%%%%%%%%%%%%%%%%%%%%%%%%%%%%%%%%%%%%%%%%%%

\newpage
\begin{center}

\mbox{\epsfig{file=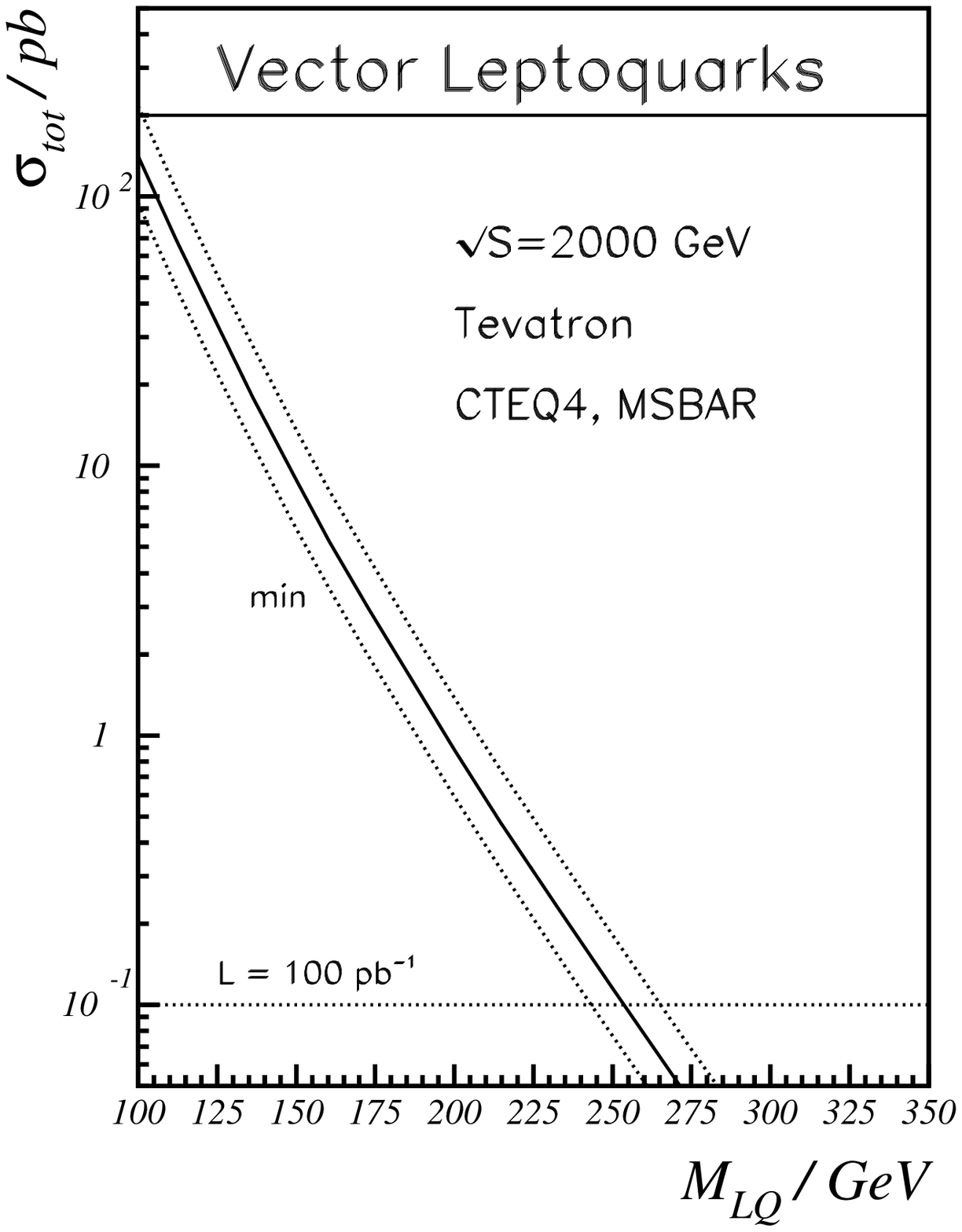,height=18cm,width=16cm}}

\vspace{2mm}
\noindent
\small
\end{center}
{\sf
Figure~10:~Integrated cross sections for vector leptoquark pair
production at Tevatron minimizing the cross section for $\kappa_G$ and
$\lambda_G$ at $\sqrt{S}~=~2.0~\TeV$. Full line~: $\mu = M_{\rm LQ}$,
dotted lines~: range of scale variation $\mu~\epsilon~[M_{\rm LQ}/2,
2 M_{\rm LQ}]$.}
\normalsize
%%%%%%%%%%%%%%%%%%%%%%%%%%%%%%%%%%%%%%%%%%%%%%%%%%%%%%%%%%%%%%%%%%%%%%%%%%

\newpage
\begin{center}

\mbox{\epsfig{file=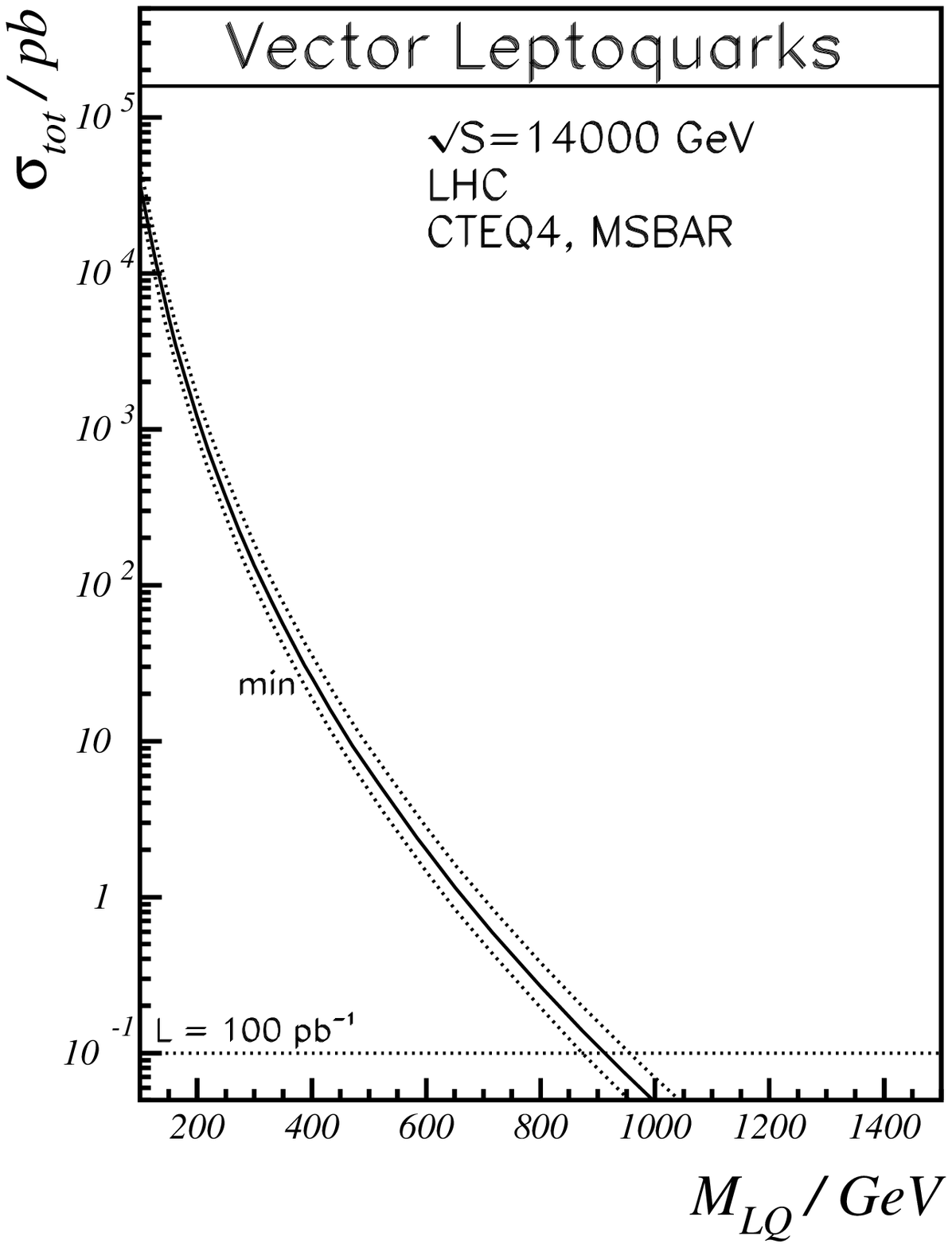,height=18cm,width=16cm}}

\vspace{2mm}
\noindent
\small
\end{center}
{\sf
Figure~11:~Integrated cross sections for vector leptoquark pair
production at LHC minimizing the cross section for $\kappa_G$ and
$\lambda_G$ at $\sqrt{S}~=~14~\TeV$. Full line~: $\mu = M_{\rm LQ}$,
dotted lines~: range of scale variation $\mu~\epsilon~[M_{\rm LQ}/2,
2 M_{\rm LQ}]$.}
\normalsize
%%%%%%%%%%%%%%%%%%%%%%%%%%%%%%%%%%%%%%%%%%%%%%%%%%%%%%%%%%%%%%%%%%%%%%%%%%
\end{document}